# Physical properties of highly uniform InGaAs Pyramidal Quantum Dots with GaAs barriers: fine structure splitting in pre-patterned substrates

L.O. Mereni, V. Dimastrodonato, G. Juska and E. Pelucchi

*Tyndall National Institute, University College Cork, Cork, Ireland*

*Corresponding author e-mail:* lorenzo.mereni@tyndall.ie

InGaAs Quantum Dots embedded in GaAs barriers, grown in inverted tetrahedral recesses of ~7 μm edge, have showed interesting characteristics in terms of uniformity and spectral narrowness of the emission. In this paper we present a study of the fine structure splitting (FSS). The investigation of about 40 pyramids revealed two main points: 1) the values of this parameter are very similar for different quantum dots proving again the uniformity of Pyramidal QDs properties, 2) there is a little chance to find a dot with natural zero splitting, but the values found (the mean being ~13 μeV) should always guarantee the capability of restoring the degeneracy with some correction technique (e.g. application of a small, easy to produce magnetic field).

**Keywords**: site-controlled quantum dots, fine structure splitting spectral purity, MOVPE, InGaAs dot, small inhomogeneous broadening.

# 1. Introduction

Semiconductor Quantum Dots (QDs) are instruments of paramount importance for investigating the properties of the low dimensional solid state. The proven capability of QDs to emit single and entangled photons [1,2] has renewed the interest for these nanostructures, especially since scientists started to look for a way to realize an appropriate photon source suitable for the nascent fields of quantum information, quantum computing and quantum key distribution. If the realization of such a "good" single photon source is a challenge on its own, an entangled photon emitter is even more delicate. Its realization with QDs is hindered by the very nature of the crystalline environment: structural asymmetries, defects, piezoelectric fields, alloy disorder [3,4,5]…each of these factors can potentially lift the degeneracy between the cascaded photon pairs, giving origin to the so called fine structure splitting (FSS) destroying entanglement.

Recent studies on site-controlled InGaAs QDs embedded in GaAs barriers grown in inverted tetrahedral recesses have showed that this system has interesting characteristics in terms of uniformity and spectral purity: the standard deviation over a broad ensemble of pyramids of the neutral exciton emission energy was found to be only 1.19 meV, while the narrowest linewidth measured (18 μeV), obtained in conditions of low excitation, set the state of the art for site-controlled systems in terms of spectral purity [6].

Given these results, it's quite natural to further investigate this kind of dots in order to understand their eligibility as high quality sources of entangled photons, especially as a system, similarly based on InGaAs Dots embedded in GaAs barriers (but with a much smaller patterning pitch, i.e. with different growth and physical properties), has very recently proven to be suitable for the scope with carefully selected dots [7]. Indeed no geometrically induced level splitting is expected *a priori* since the $C_{3v}$ symmetry of the tetrahedral recess (fig.4-inset) as a result of the (111) crystallographic symmetry does not exhibit any preferential axis. This has lead a number of authors [8,9,10] to suggest QDs grown on (111) surfaces as ideal for the production of entangled photons sources.

We present here the results of the study on about 40 QDs in one of our best samples [6] in order to estimate the range of values of the fine structure splitting (FSS) characterizing the system. The main feature emerging from the study is a high uniformity and a small value of this parameter especially if compared to Stranski-Krastanov QDs: the average value of the splitting was found to be 12.9 µeV. Further work will be needed to elucidate the origin of such splitting in order to improve our control and to reproducibly obtain nearly zero FSS.

## 2. Results and discussion

The epitaxial growth of the layers occurs by metallorganic vapour phase epitaxy (MOVPE) on GaAs (111)B ±0.1° substrate pre-patterned with a regular matrix of 7.5µm-pitch inverted tetrahedral recesses. As result of the selective etching process, the pyramids expose sharp lateral facets with crystallographic orientation (111)A[11]. The temperature during the growth of the inner barriers and the QD itself is kept around 730 ºC (thermocouple reading), the V/III ratio in the range 600-800. The QD layer is composed by $In_{0.25}Ga_{0.75}As$ nominally 0.5 nm thick, the barriers by pure GaAs.

The measurements of the fine structure splitting were taken at 10K by high resolution (5 µeV) polarization resolving spectroscopy (basically by rotation of a half waveplate preceding a fixed polarizer in the signal path). The sample was illuminated by a continuous red diode laser at 656 nm; light was collected with a 100X microscope objective and sent to 1 metre long monochromator equipped with a 2048 pixels CCD array . The measurements were taken with the sample in apex down geometry [6]. Therefore relatively high pump powers were to be used during the measurements with the consequent line broadening due to excess carrier generation. In these conditions the average linewidth of the exciton line was around 100 μeV as explained in ref [6].

In Fig.1 is shown a schematic representation of the effects of FSS [2]: in presence of splitting the otherwise circularly polarized and indistinguishable photons generated by the recombination of neutral excitons (X) in the dot, become a linearly-cross polarized doublet as the two possible transitions are no longer isoenergetic. Being formed two decay paths for the biexciton (XX), also its peak will become a doublet characterized by the same splitting as its exciton counterpart. The polarizer allows the discrimination of the two linear polarizations by letting through only photons with a polarization parallel to the extraordinary axis while those with orthogonal polarization are rejected.

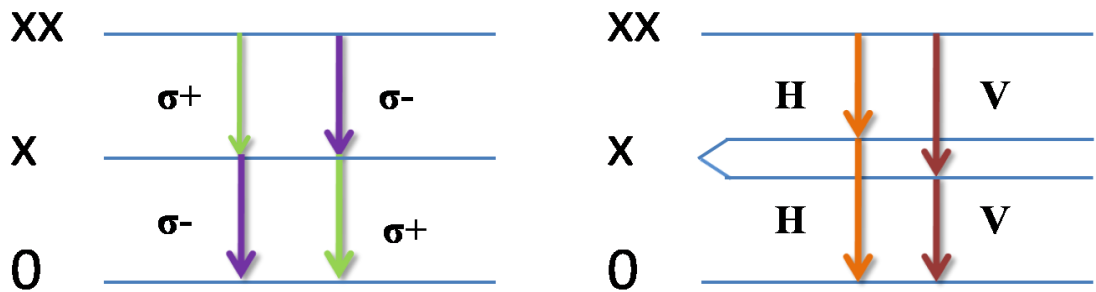

*Figure 1*

In Fig. 2 two representative spectra of the same single QD are taken at orthogonal angles of the half-waveplate and shown after fitting the experimental data with Lorentzian lineshapes. The spectra are then superimposed to point out the splitting.

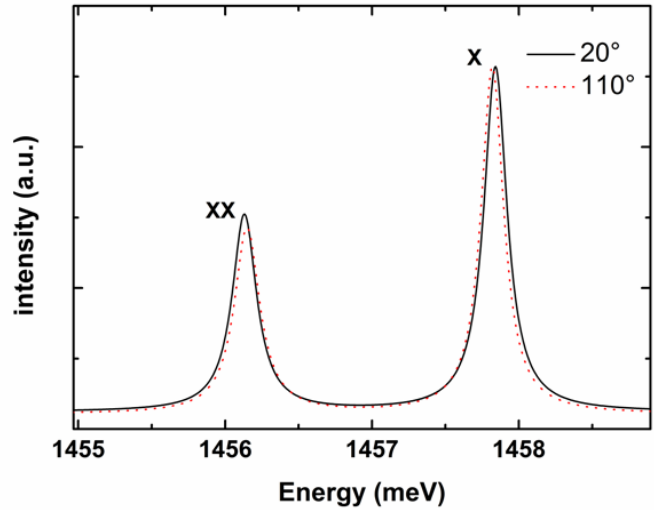

*Figure 2*

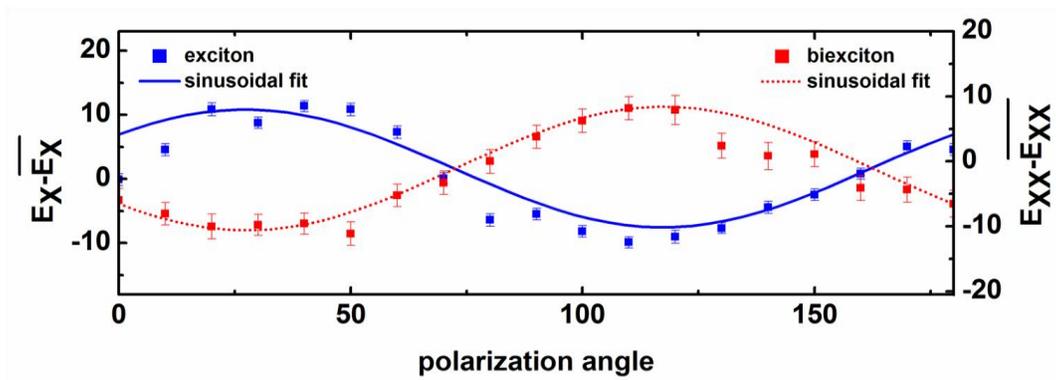

*Figure 3*

The spectra of almost all the pyramids were dominated by the X and the XX lines. The average exciton emission energy is 1457.6±2.4 meV. The binding energy between the biexciton and the exciton is positive and around 1.5 meV. Despite the $C_{3v}$ symmetry of the dots, we measured a FSS in all the dots: in the case of Fig.2 the FSS was ~18 μeV.

Fig.3 shows the X (solid) and XX (dotted) peak positions offset by their average value as a function of the polarization angle: the expected counterphase behaviour is evident.

Similar measurements were done on almost 40 pyramids. The FSS values were ranging from 5 μeV to 21 μeV. Assuming a Gaussian distribution we found that the mean value is 12.9 μeV and the standard deviation is only 3.8 μeV (Fig.4).

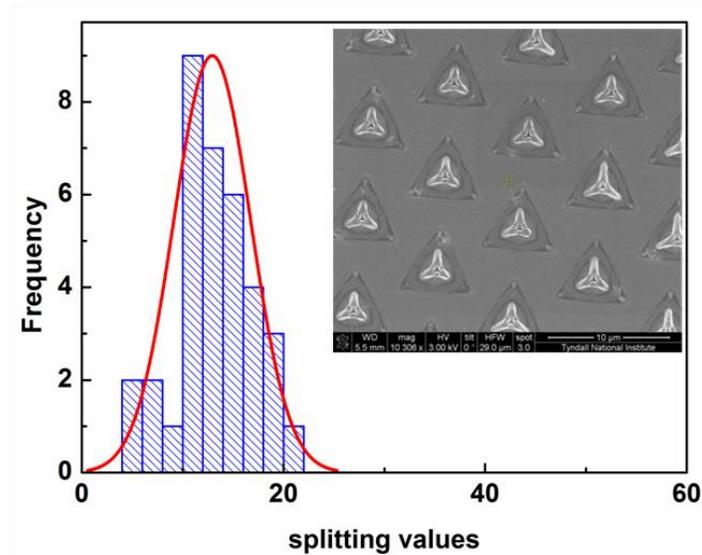

*Figure 4*

Unfortunately the zero value lays more than three σ away from the centre of the distribution meaning that the probability of finding a dot "naturally" suitable for entangled photon emission is very low. The value, though, is sufficiently small (even if compared with other pyramidal QD systems [12]) to assure the feasibility of corrections (using a magnetic field [13] or applying time gating techniques [14]). On the other hand the small range of values found is another proof of the uniformity of the pyramidal dots with GaAs barrier and its versatility when compared to Stranski-Krastanow QDs and/or other site controlled systems.

The origin of the splitting in these dots remains to be understood, even though some hypothesis can be formulated: it is possible that we are observing the effect induced by the inevitable alloy disorder due to the Indium random distribution inside the InGaAs alloy of the dot, the effect of which, while not affecting bulk materials, could not be averaged given the nanoscopic dimensions of the dot [4]. Another hypothesis is that the apparently perfect tetrahedral recess is affected by lithographically induced imperfections that are not corrected during the growth and this could be the cause of the unexpected loss of symmetry. Further studies involving the phases of the FSS and measurements of dots with different alloy compositions should help clarifying some of these points as preliminary measurements have shown a preferential orientation of the low energy axis of the exciton towards one in particular of the three equivalent (110) directions, i.e. towards one of the pyramid in-plane apexes, which could suggest a lithographic origin of the residual dot asymmetry.

**Acknowledgments**

This research was enabled by the Irish Higher Education Authority Program for Research in Third Level Institutions (2007-2011) via the INSPIRE programme, and by Science Foundation Ireland under grants 05/IN.1/I25. We are grateful to K. Thomas for his support with the MOVPE system and to Robert J. Young for the hints on the FSS measurements.

**References**

[1] P. Michler et al., Science **290**, 2282 (2000)
[2] A. J. Shields et al., Nature Photonics **1**, 215-223 (2007)
[3] M. Abbarchi et al., Phys. Rev. B, **78**, 125321 (2008)
[4] V. Mlinar and A. Zunger, Phys. Rev. B **79**, 115416 (2009)
[5] J. D. Plumhof et al., Phys. Rev. B **81**, 121309 (2010)
[6] L. O. Mereni et al., App. Phys. Lett. **94**, 223121 (2009)
[7] A. Mohan et al., Nature Photonics **DOI 10.1038** (2010)

[8] A. Schliwa et al., Phys. Rev. B **80**, 161307 (2009)
[9] V. Troncale et al., J. Appl. Phys. **101**,081703 (2007)
[10] R. Singh et al., Phys. Rev. Lett. **103**, 063601 (2009)
[11] E. Pelucchi et al., Nano Lett. **7**, 1282 (2007)
[12] D. Y. Oberli et al., Phys. Rev. B **80**, 165312 (2009)
[13] Stevenson et al., Phys. Rev. B **73**, 33306 (2006)
[14] Stevenson et al., Phys. Rev. Lett. **101**, 170501 (2008)

# Figure captions

**Figure 1** (*colour online*): Decay paths for the exciton (X) and biexciton (XX) in case of degenerate levels (left) and split levels (right). In the first case the photon emitted are circular, in the second the mixing of states produced by asymmetries gives rise to linearly polarized photons.

**Figure 2** (*colour online*): Lorentzian fit of single dot representative exciton (higher energy) and biexciton emission for two different polarization angles (20° and 110°). The measured value of the fine structure splitting in this case is 18 µeV. The estimated pump power on the sample was~2 µW.

**Figure 3** (*colour online*): positions of exciton (solid line) and biexciton (dotted line) peaks plotted against the polarization angles for the same dot that appears in Fig.2. The average position of the biexciton (exciton) has been offset by its average.

**Figure 4** (*colour online*): Distribution of values of the FSS. Fitting the distribution with a Gaussian gives a mean value of 12.9 µeV and a standard deviation of 3.8 µeV. The inset shows a SEM image of the ordered matrix of Quantum Dots. The distance between two pyramids is 7.5 µm.